# Performance Prediction of InP/GaAsSb Double Heterojunction Bipolar Transistors for THz applications

Xin Wen,[1, a)] Akshay Arabhavi,[2] Wei Quan,[2] Olivier Ostinelli,[2] Chhandak Mukherjee,[3] Marina Deng,[3] Sébastien Frégonèse,[3] Thomas Zimmer,[3] Cristell Maneux,[3] Colombo R. Bolognesi,[2] and Mathieu Luisier[1]

[1)] *Integrated Systems Laboratory, Department of Electrical Engineering and Information Technology, ETH Zürich, Gloriastrasse 35, 8092 Zürich, Switzerland.*
[2)] *Millimeter-Wave Electronics Laboratory, Department of Electrical Engineering and Information Technology, ETH Zürich, Gloriastrasse 35, 8092 Zürich, Switzerland.*
[3)] *IMS Laboratory, University of Bordeaux, 351, Cours de la Libération, 33405 Talence, France.*

*Abstract*—The intrinsic performance of "type-II" InP/GaAsSb double heterojunction bipolar transistors (DHBTs) towards and beyond THz is predicted and analyzed based on a multi-scale technology computer aided design (TCAD) modeling platform calibrated against experimental measurements. Two-dimensional hydrodynamic simulations are combined with 1-D full-band, atomistic quantum transport calculations to shed light on future DHBT generations whose dimensions are decreased step-by-step, starting from the current device configuration. Simulations predict that a peak transit frequency $f_{T,peak}$ of around 1.6 THz could be reached in aggressively scaled type-II DHBTs with a total thickness of 256 nm and an emitter width $W_E$ of 37.5 nm. The corresponding breakdown voltage $BV_{CEO}$ is estimated to be 2.2 V. The investigations are put in perspective with two DHBT performance limiting factors, self-heating and breakdown characteristics.

## I. INTRODUCTION

The InP double heterojunction bipolar transistor (DHBT) technology has become a promising candidate to develop terahertz (THz) circuits and systems for both industrial and research applications. By providing impressive "cut-off frequency times breakdown voltage" ($f_T \times BV_{CEO}$) products and high power handling capabilities at room temperature, the InP DHBT platform has established itself as a very competitive approach to reach THz operating frequencies [1-3]. Among its members, type-II InP/GaAsSb DHBTs are particularly attractive as they do not exhibit any energy barrier at the base-collector heterojunction, contrary to type-I InP/InGaAs components. This feature allows for a simplified base-collector design and a higher breakdown voltage at a given cut-off frequency [2]. Recently, type-II InP/GaInAsSb DHBTs with a quaternary base have been fabricated, achieving better transient performance than their ternary counterparts because of the larger Γ-L valley separation of GaInAsSb [4].

TCAD-based roadmaps towards THz applications have already been developed for different heterojunction bipolar transistor (HBT) technologies [5, 6]. The electrical characteristics in these studies were evaluated based on classical drift-diffusion (DD) or hydrodynamic (HD) models calibrated with a full-band Boltzmann Transport Equation solver [7] indirectly accounting for tunneling currents. As experimental research efforts progress towards THz frequencies, the quantum mechanical effects play an increasingly critical role due to the reduction of the device dimensions along all directions and the combination of always more complex materials. A fully quantum mechanical approach allows to naturally and automatically integrate all related effects. In this work, the transport characteristics of type-II InP/GaAsSb DHBTs are studied theoretically based on the scaling prescriptions of Ref. [8], including electron-phonon interactions and breakdown effects.

To calculate the electronic transport properties of the considered devices, a multi-scale TCAD environment relying on a 2-D hydrodynamic (HD) simulator and a ballistic 1-D full-band, atomistic quantum transport (QT) solver has been employed [9]. Such an approach offers a good compromise between accuracy (bandstructure and quantum mechanical effects are included) and computational efficiency (the electrostatics come from the classical calculations). Despite these attractive features, the proposed methodology suffers from the absence of electron-phonon interactions and possibly self-heating effects. This has so far limited our investigations to the low-injection regime up to the peak of $f_T$. Furthermore, without this dissipative scattering source the transit times through the DHBT might be severely underestimated over the entire injection range. Consequently, we go beyond ballistic investigations of the original and proposed down-scaled transistors and present simulation results accounting for electron-phonon interactions via the non-equilibrium Green's function (NEGF) formalism. The importance of these effects will be discussed for each DHBT generation. This addition is expected to provide a more accurate description of the DHBT behavior. Moreover, two models have been developed to account for self-heating, one where the phonon population is driven out of equilibrium [10], and one where the dissipated power is used as source term in the classical Fourier equation [11].

The epitaxial structure of an experimentally fabricated type-II InP/GaAsSb DHBT with an emitter area $A_E = 0.3 \times 7.5$ μm$^2$ serves as starting point for our analysis. The geometries of three subsequent device generations have then been derived from this

a) Author to whom correspondence should be addressed: wenx@iis.ee.ethz.ch.

initial structure by scaling down its dimensions both vertically and laterally according to the prescriptions of [8], while adjusting the doping concentrations. The intrinsic transport characteristics of these four device generations have then been simulated with and without electron-phonon scattering and compared to each other. It has been found that a $f_{T,peak}$ of around 1.6 THz, with a breakdown voltage $BV_{CEO}$ of 2.2 V could be reached for aggressively scaled type-II DHBTs. Due to the excellent transport properties of III-V semiconductors, it has been observed that turning on electron-phonon scattering only leads to a reduction of the electronic current by 10%, as compared to the ballistic case, which does not significantly increase the lattice temperature at the center of the device ($\Delta T < 10$ K). It should be noted that the inclusion of the two lateral dimensions, the collector current crowding effect, and of the parasitic elements might raise the lattice temperature. Due to the high computational burden associated with such effects, they could not be taken into account here. To compensate for the potentially underestimated lattice temperature, device simulations at 400 K have been performed as well, with little influence observed on the cut-off frequencies.

The paper is organized as follows: in Sec. II, the electronic transport properties of the four proposed device generations are simulated utilizing the calibrated multi-scale TCAD framework of Ref. [9]. In Sec. III, the influence of electron-phonon scattering and self-heating is investigated through two different QT approaches allowing for the extraction of an effective lattice temperature. In Sec. IV, the breakdown characteristics of the vertically scaled structures are analyzed using the ballistic simulation workflow presented in Sec. II. The ballistic assumption is justified by the fact that the introduction of phonon does not significantly affect the electrostatics, which is responsible for the onset of the breakdown mechanism.

## II. BALLISTIC TRANSPORT SIMULATIONS

The calibrated TCAD model described in [9] is applied to study the THz properties of DHBTs as their dimensions are scaled down. In this approach, 2-D DC device simulations based on the hydrodynamic model of Sentaurus-Device [12] are coupled to 1-D ballistic electronic transport calculations performed in the NEGF formalism. The latter calculations rely on an empirical tight-binding (TB) basis set parameterized for InP/GaAsSb heterojunctions [13]. It has been shown in [9] that the proposed modeling framework is capable of providing good qualitative and reasonable quantitative agreements with experimental measurements of both type-I (InGaAs/InP) and type-II (GaAsSb/InP) DHBT technologies.

In this work, the epitaxial structure of the type-II DHBT described in [9] defines our initial device structure labeled G0. This choice is based on the fact that G0 represents one of the most advanced high-speed transistors to date [2, 14, 15]: it is indeed capable of simultaneously delivering high $f_T$ and $f_{MAX}$. Starting from there the scaling prescriptions of [8] are adopted. The basic principle consists of increasing the bandwidth of a given transistor by a factor $\gamma$ through a decrease of its emitter ($T_E$) and base ($T_B$) thicknesses by a factor slightly higher than $\gamma^{1/2}$ and a simultaneous reduction of the collector thickness $T_C$ by $\gamma$. At the same time, the emitter area $A_E$ must be scaled down by $\gamma^2$ for an optimally balanced design between the cut-off frequency ($f_T$), maximum frequency ($f_{MAX}$), and thermal budget [8].

Based on the measured transient characteristics of G0 ($f_{T,peak}$ = 0.46 THz) and our objective of an ultimate DHBT delivering a peak cut-off frequency $f_{T,peak}$ above 1 THz, a roadmap containing three generations of transistors, G1 to G3, has been defined. For each of them, an increase of about 30% in terms of the peak cut-off frequency is set as target. To realize that, the vertical dimensions of the emitter and the base ($T_E$ and $T_B$) are shrunk by 15% and the thickness of the collector ($T_C$) by 30%. Besides, the emitter and collector widths of the device ($W_E$ and $W_C$) are decreased by 50%. At the same time, the doping concentrations in the emitter and collector regions are increased accordingly, whereas the graded base doping with an average value of $8.5 \times 10^{19}$ cm$^{-3}$ is kept unchanged to ensure a suitable electrostatics. Abrupt doping profile variations from one layer to the other is assumed, i.e. no diffusion of dopants has been taken into account. We do not expect this approximation to play a critical role as the experimental data for G0 could be reproduced in this doping condition. The definition of two intermediate generations between G0 and G3 is justified by the fact that technology developments will be needed at each step to process the device geometries and reach the desired doping concentrations.

The constructed DHBT structures are then fed into our calibrated TCAD framework to derive the corresponding DC and AC characteristics. During this procedure, the material parameters that have been calibrated for G0 are kept unchanged. All simulations are done at room temperature (T = 300 K) and in the ballistic limit of transport for the quantum mechanical part. Orthorhombic unit cells of size of $L_x = L_y = L_z$ = 5.868 Å that are repeated along the transport direction (x) are used in this case. A homogeneous energy grid with a spacing of 2 meV between adjacent points is utilized in all QT simulations.

The conduction band edge and the corresponding doping profile at $V_{CE} = 1$ V, $V_{BE} = 0.94$ V, and a collector density $J_C$ = 49.4 mA/µm$^2$ are plotted in Fig. 1 for the final structure G3. The base thickness $T_B$, after aggressive vertical scaling, measures 12.3 nm only, with a collector thickness $T_C$ of 43.8 nm. The conduction band offset at the emitter-base interface is equal to $\Delta E_C = 139$ meV, with the InP value as a reference. The increased doping concentrations in the emitter and collector layers lead to a larger bandgap narrowing (BGN) than at G0, thus resulting in larger conduction band discontinuities at the E-B and B-C interfaces. The differences between the Γ and L valleys are $\Delta E_{\Gamma L} = 196$ meV in the base [16] and 590 meV in the InP regions. These values remain the same for all device generations.

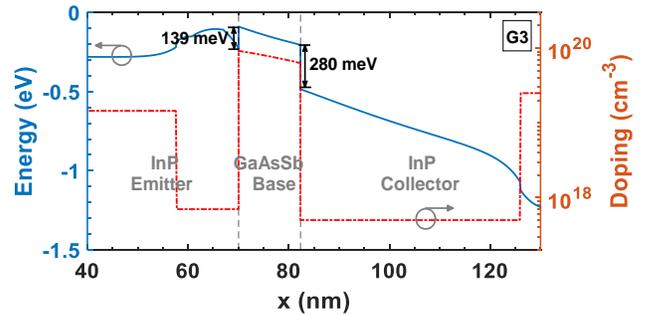

FIG. 1: Conduction band diagram (as obtained from S-Device, left y-axis) and the corresponding doping concentration (right y-axis) of the type-II InP/GaAsSb DHBT G3 structure at applied biases $V_{CE} = 1$ V and $V_{BE} = 0.94$ V, with a collector current density $J_C = 49.4$ mA/µm$^2$. The variable x is the distance from the emitter contact.

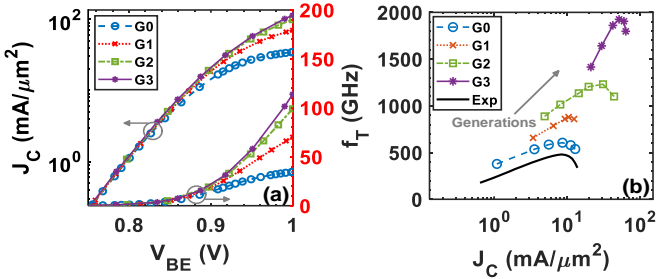

FIG. 2: Simulated DC and AC characteristics of all InP/GaAsSb DHBT generations. (a) 2-D HD simulated collector current density $J_C$ versus base-emitter voltage $V_{BE}$ at $V_{BC} = 0$ V plotted on a logarithmic (left axis) and linear (right axis) scale. (b) Cut-off frequency $f_T$ as a function of $J_C$ at $V_{CE} = 1$ V computed from 1-D QT calculations. The solid line without symbols represents the experimental measurements for G0 [4].

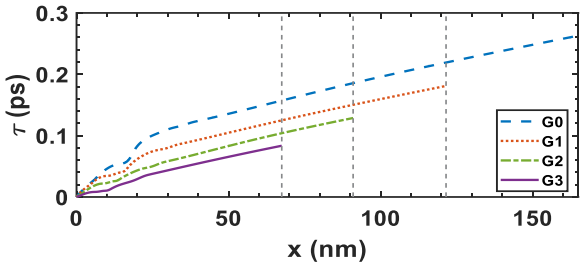

FIG. 3: Accumulated intrinsic transit times extracted from 1-D QT simulations along the emitter-collector axis corresponding to the peak cut-off frequencies $f_{T,peak}$ at $V_{CE} = 1$ V for all InP/GaAsSb DHBT generations. The variable x is the distance from the top of the active emitter region. The vertical dashed lines indicate the end of the active collector regions of each device generation, respectively.

The emitter-collector transit time is obtained from the following equation:

$$\tau(x) = q \int_0^x \frac{\Delta n}{\Delta J_c} dx \quad (1),$$

where the derivative of the electron concentration $n$ with respect to the collector current density $J_C$ are integrated along a 1-D line across the active E-B-C junction [17]. The variations of $n$ and $J_C$ are computed from two QT simulations with slightly different $V_{BE}$ ($\Delta V_{BE} = 1$ meV in our studies). The transit time in the quasi-neutral region of the emitter is multiplied by an additional factor of 2/3 to compensate for the injected minority charges that are not reclaimable through this junction, i.e. those that are lost, for example through recombination processes [18]. The 1-D QT simulated transit frequency $f_T$ is then calculated from a simple expression

$$f_T = \frac{1}{2\pi\tau} \quad (2),$$

where $\tau$ can be computed from Eq. (1) in the intrinsic region of the transistor. Note that the contributions from parasitic elements are not taken into account in Eq. (2). Figure 2(a) summarizes the collector current densities $J_C$ versus $V_{BE}$ for each generation, when $V_{BC} = 0$ V. From G0 to G3, the decreased device area and the increased electric field lead to enhanced $J_C$ values. Because of the growing collector doping concentrations, the onset of the Kirk effect shifts to slightly higher $V_{BE}$ values when one moves from G0 to G3. The "$f_T$ vs. $J_C$" curves at $V_{CE} = 1$ V, as calculated from the ballistic QT simulations, are presented in Fig. 2(b), together with the experimental results for G0 as a reference. In terms of the experimental results, iterative de-embedding method was used after measuring the S-parameters of the transistors to subtract the parasitics associated with the device probing pads [19]. The simulated $f_T$ is overestimated by about 20% in the middle- to high-injection regimes due to the absence of non-idealities, i.e., electron-phonon interactions, lateral extensions of the transistor, and parasitic elements. All these factors could lead to further $f_T$ reductions as compared to 1-D ballistic simulations. In addition, the observed larger discrepancy of around 35% in the low-injection regime ($J_C < 3$ mA/μm²) can be attributed to the possible underestimation of the conduction band offset at the E-B interface. More specifically, during the calibration process of G0, it was observed that an increase in the conduction band offset at the E-B interface leads to a more significant reduction of the simulated $f_T$ in the low-injection regime than in the middle- and high-injection ones. However, since we are more interested in the middle- to high-injection regions in this study and considering the fact that the obtained qualitative and quantitative agreements are already satisfying, no further fine tuning was carried out.

As expected, $f_{T,peak}$ increases by about 30% between two consecutive generations, with a final value of 1.9 THz in G3. It should however be noted that the collector current densities at $f_{T,peak}$ increase with scaling, which is expected to favor self-heating effects. At very high current injection levels, it is believed that $f_T$ diminishes due to the formation of local hot spots that reduces the carrier mobility [9]. Accounting for such phenomena requires the inclusion of electron-phonon interactions in the QT simulations, as discussed in Sec. III.

The contributions to the intrinsic transit times from the emitter, base, and collector regions at $f_{T,peak}$ are presented in Fig. 3 for all considered device geometries. Typically, the $f_T$ improvements can be attributed to the vertical scaling of the devices. The observed sharp increase of the transit time in the E-B depletion region of the G0 DHBT (at around x ~ 20 nm) becomes less important in the next generations G1-G3. This rapid increase is caused by the E-B conduction band offset and is partly compensated by the scaled dimensions in G1 to G3.

## III. DISSIPATIVE TRANSPORT SIMULATIONS

With continuous scaling of device dimensions, management of thermal budget in DHBTs is critical, especially in InP components [20], [21]. As the $f_{T,peak}$ values around which modern bipolar transistors operate are pushed toward higher current densities, self-heating effects become increasingly important. Consequently, to refine our analysis, dissipative transport simulations have been implemented utilizing two different models. In the first one (model 1), the electron and phonon populations are self-consistently coupled to each other via scattering self-energies within the NEGF formalism, as explained in [10] and Appendix A. Both current and energy conservations are ensured. However, the limitation of model 1 comes from the fact that it does not include polar-optical phonon (POP) scattering, which is essential in III-V semiconductors [22]. As an alternative, a second model (model 2) has been implemented, where POP is taken into account, but the temperature is treated classically through Fourier's equation [11], as described in Appendix B. In both cases, the same HD electrostatic potentials as in the ballistic case are used. Once self-consistency between the Green's Functions and scattering self-energies is achieved, the transit times and the cut-off frequencies are calculated as in Sec. II.

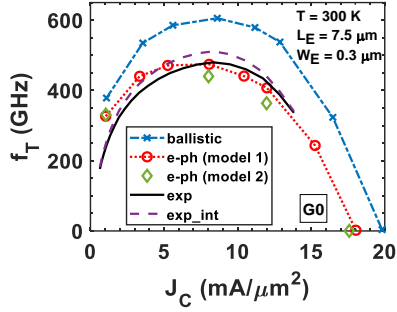

FIG 4: One-dimensional QT simulated "$f_T$ vs. $J_C$" curves for G0 with (dotted line with circles) and without (dash-dotted line with crosses) electron-phonon scattering at T = 300 K with model 1 and model 2 (diamonds). The experimental results with (solid line) and without (dashed line) parasitic elements are provided as a reference. All data are extracted at $V_{CE}$ = 1 V.

The electron-phonon coupling strength of model 1 and 2 was calibrated at T = 300 K against experimental AC measurements for G0 and was kept constant in all simulations. This assumption is justified by the fact that the electron-phonon scattering strength mostly depends on the material systems and the device lateral dimensions. As the same combination of materials will be used from G0 to G3 and because phonon confinement has a limited influence in structures laterally extending over tens of nanometers, constant electron-phonon interactions appear as a reasonable choice. Figure 4 compares the simulated $f_T$ with and without electron-phonon scattering with measured data. Due to the 1-D nature of the QT simulation domain, parasitic elements are not accounted for. Hence, the experimental $f_T$ is corrected to eliminate these effects and becomes $f_{T,int}$:

$$f_{T,int} = \frac{1}{\frac{1}{f_T} - 2\pi C_{BC}(R_{CX}+R_E)} \quad (3),$$

where $R_{CX}$ represents the extrinsic collector resistance, $R_E$ the emitter resistance, and $C_{BC}$ the junction capacitance of the B-C region. In G0, they are equal to 3 Ω, 1.7 Ω, and 5 fF, respectively. These values were extracted from the experimental measurements. The "$f_T$ vs. $J_C$" curves computed with model 1 and 2 agree well with the experimental data over the entire injection range. The slight overestimation in the low-injection regime might be due to the underestimation of the conduction band offset at the E-B interface, which leads to higher current densities.

As compared to the ballistic case, dissipative transport induces a reduction of the collector current density and of the associated cut-off frequency for G0, as illustrated in Fig. 5. Generally, the reductions in current density and cut-off frequency are more important at higher collector current densities. This can be explained by the fact that when $J_C$ increases, higher energy states become populated by electrons. These electrons can emit phonons more easily, which enhances the backscattering probability. The ratio between the dissipative and ballistic cut-off frequencies, $f_{Ts}/f_{Tb}$ then reaches a minimum of around 50% at $J_{Cb}$ = 20 mA/μm². Similarly, the reduction in current density caused by electron-phonon scattering is shown in Fig. 5(b) as a function of $V_{BE}$. It does not surpass 10%, even in the high-injection regime. This value is much lower than in Si [23].

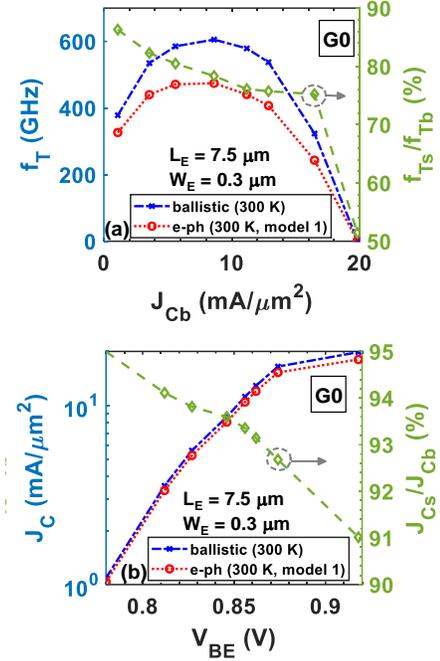

FIG 5: (a) 1-D QT simulated "$f_T$ vs. $J_{Cb}$" characteristic with (dotted line with circles) and without (dash-dotted line with crosses) electron-phonon scattering for G0 (left axis) at T = 300 K. Model 1 was used for that purpose. The corresponding $f_{Ts}/f_{Tb}$ ratio is shown on the right axis. Here, $J_{Cb}$ is the ballistic collector current density, $f_{Ts}$ the cut-off frequency with electron-phonon scattering, and $f_{Tb}$ the same quantity in the ballistic limit at $V_{CE}$ = 1 V. (b) 1-D QT simulated collector current density as a function of $V_{BE}$ with (dotted line with circles) and without (dash-dotted line with crosses) electron-phonon scattering for G0 (left axis) and the corresponding $J_{Cs}/J_{Cb}$ ratio (right axis), all at T = 300 K and $V_{CE}$ = 1 V, as obtained with model 1. $J_{Cs}$ represents the collector current density with electron-phonon scattering.

In ballistic simulations, the $f_T$ reduction after reaching its peak value can be mainly attributed to the well-known Kirk effect, which is included in our simulations by reading the electrostatic potential from the 2-D HD model. The other important factors that lead to the experimentally observed $f_T$ decrease in the high-injection regime should be accounted for as well, which include self-heating and the formation of hot spots that locally decrease the electron mobility [9]. To quantify the influence of self-heating, the effective lattice temperature of the G0 DHBT, as extracted with model 1 and 2, is plotted in Fig. 6. Both models deliver temperatures that are very close to each other and that do not reach high values, contrary to what has been theoretically predicted for SiGe bipolar transistors [24]. Initially, based on the results of model 1 only, we believed that the moderate temperature increase was due to the absence of polar optical phonon scattering (see Appendix A). However, model 2, which accounts for this effect in a simplified way, essentially leads to the same results (see Fig. 6(b)), suggesting that something else influences the temperature behavior.

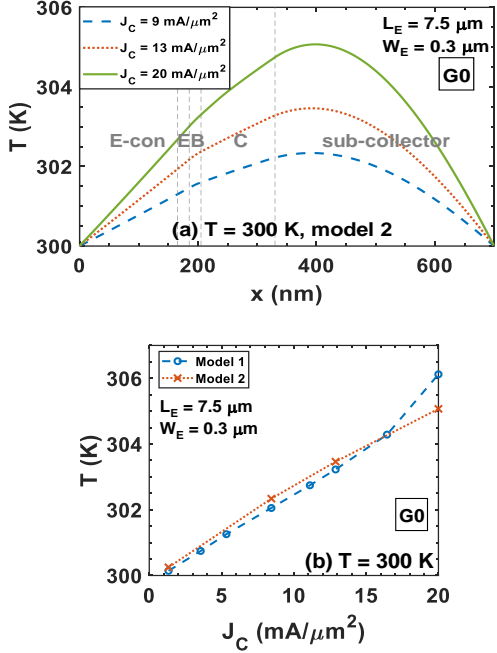

FIG 6: (a) 1-D QT simulated lattice temperature profile with electron-phonon interactions across the center of the G0 transistor at $J_C$ = 9, 13 and 20 mA/µm², $V_{CE}$ = 1 V, as computed with model 2, assuming a surrounding temperature T = 300 K. The variable x is the distance from the emitter contact. (b) Maximum of the lattice temperature as a function of $J_C$ for model 1 (dashed line with circles) and model 2 (dotted line with crosses), as obtained from 1-D QT calculations.

To shed light on this issue, we extracted the amount of electrical power dissipated inside the simulation domain, $P_{diss,in}$ This quantity can be obtained from the energy current ($J_{CE}$, unit: W/m²), as defined in Appendix B, by considering the difference between the emitter and collector values. Normally, it should be equal to $P_{diss,tot} = V_{CE} \times I_C$. However, in our simulations of the G0 DHBT, $P_{diss,in}$ does not exceed 12% of $P_{diss,tot}$ at $I_C$ = 45 mA, despite the length of the 1-D domain (700 nm). This means that at $V_{CE}$ = 1 V, electrons lose less than 0.12 eV of their energy, instead of 1 eV. Such a loss corresponds to the emission of roughly four optical phonons, according to the parameters listed in Appendix B. From the distribution of $J_{CE}$, it appears that the power dissipation mainly occurs in the collector and sub-collector regions, which measure about 500 nm in total. It follows that the electron mean free path for scattering, $L_{mfp}$, in the G0 DHBT, can be estimated to be ~125 nm. In SiGe, $L_{mfp}$ is much shorter, thus leading to higher temperatures [24].

Since the power dissipation $P_{diss,in}$ inside the simulation domain of the G0 DHBT does not exceed 12% of the total dissipation calculated as $V_{CE} \times I_C$, the total length of the collector and sub-collector region $L_{Ctot}$ needs to be about 1/0.12=8.3 times longer than the dimension of the simulated structure ($L_{Ctot0}$ = 0.5 µm) to completely dissipate the electrical power and convert it to heat. Therefore, a $L_{Ctot}$ of 0.5×8.3 ≈ 4.2 µm is required. On one hand, such dimensions are greater than what can be actually simulated. On the other hand, they are also larger than the actual experimental device. Hence, it can be expected that part of the power dissipation happens in the metallic contact of the emitter and may not contribute to the self-heating of the active device area. It is assumed that the amount of heat dissipated in the contacts does not contribute to an increase of the device temperature and is therefore omitted from the 1-D QT simulations.

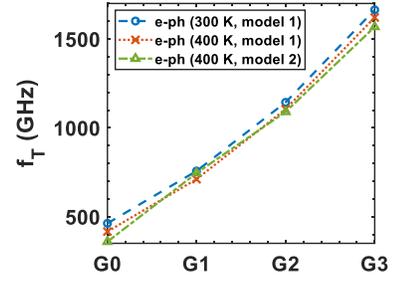

FIG 7: 1-D QT simulated $f_{T,peak}$ for all device generations, G0 to G3, as calculated with model 1 at T = 300 K (dashed line with circles), with model 1 at T = 400 K (dotted line with crosses), and with model 2 at T = 400 K (dash-dotted line with triangles). All simulations are performed at $V_{CE}$ = 1 V. The dissipative current densities that produce $f_{T,peak}$ are 8.1 mA/µm², 9.9 mA/µm², 29.6 mA/µm², and 49.5 mA/µm² for G0, G1, G2, and G3 at 300 K with model 1, respectively.

The fact that our simulation results agree well with experiments in Fig. 4 tends to indicate that self-heating might not play a detrimental role in type-II DHBTs. The inclusion of neglected contributing factors, e.g. the lateral device extensions and current crowding effects could change the situation and push up the temperature increase, ΔT. For example, slightly higher values are proposed in [25], and much higher ones are reported in [26] and [27].

To go one step further, we estimated the thermal resistance $R_{th}$ of a structure similar to G0 based on the temperature-dependent measurements of the collector current $I_C$ and following the procedure of Ref. [28]. This data revealed a $R_{th}$ of 4000 K/W at $I_C$ = 22.5 mA and $V_{CE}$ = 1 V. If the entire electrical power is assumed to be dissipated within the active DHBT region, the determined $R_{th}$ leads to a temperature increase ΔT = $R_{th} \times P_{diss,tot}$ = 90 K at $I_C$ = 22.5 mA and $V_{CE}$ = 1 V. To take this ΔT into account, we calculated $f_{T,peak}$ for each DHBT generation at two different internal temperatures, 300 and 400 K, with electron-phonon interactions, as implemented in model 1 and 2. Results can be found in Fig. 7.

It can be seen there that $f_{T,peak}$ increases by roughly 30% at each generation, reaching 1.66 (1.57-1.62) THz at G3, if the lattice temperature of the active region is equal to 300 (400) K. Compared to the ballistic case, the inclusion of electron-phonon interactions leads to a reduction in $J_C$ for all device generations. This reduction decreases from 6.4 to 3.7% at 300 K with model 1, as the transistor evolves from G0 to G3. The diminishing difference between the ballistic and dissipative results can be explained by the fact that the impact induced by the down-scaled device dimensions outperforms that by the increasing collector current densities. Overall, a ΔT of 100 K only marginally affects $f_{T,peak}$, regardless of the DHBT generation, but the general trend remains exactly the same in both cases. For G0, the simulated $f_{T,peak}$ with electron-phonon scattering from model 1 decreases by 47 GHz when T is increased from 300 to 400 K. As a reference, the experimental peak $f_T$ reported in [4] reduces by ~40 GHz when the temperature is increased from 200 to 300 K. Besides, the measured reduction in $f_{T,peak}$ is around 60 GHz when the temperature is increased from 300 to 400 K at $V_{CE}$ = 1 V for a device with similar epitaxial structure [29]. Therefore, the simulated variation in $f_{T,peak}$ with increased temperature stays in a meaningful range. It can be concluded from these results that increasing the temperature of the active DHBT region does not significantly deteriorate its AC characteristics. However, it cannot be excluded that temperature-induced damages might take place at the semiconductor-contact interface, if a large portion of the electrical power is dissipated in this area or in the sub-collector region. The missing 2-D effect induced by the lateral extension of the transistors could possibly reduce the simulated $f_{T,peak}$. The omission of these mechanisms could lead to an underestimation of the dependence of $f_T$ on the temperature, especially for the future device generations G1-G3 with higher $J_C$.

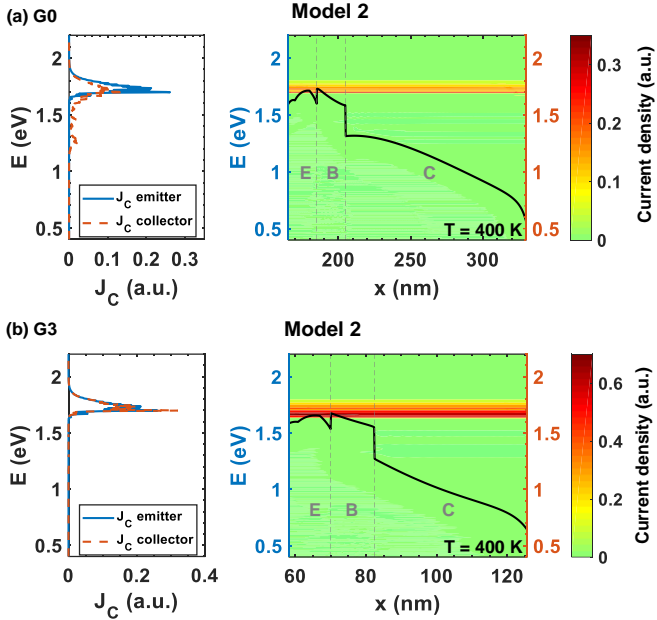

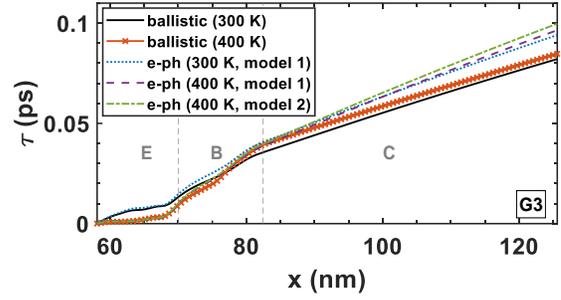

FIG 9: 1-D QT computed accumulated transit times extracted with and without electron-phonon scattering at $f_{T,peak}$ ($V_{CE}$ = 1 V), T = 300 and 400 K for the G3 DHBT. The variable x is the distance from the emitter contact.

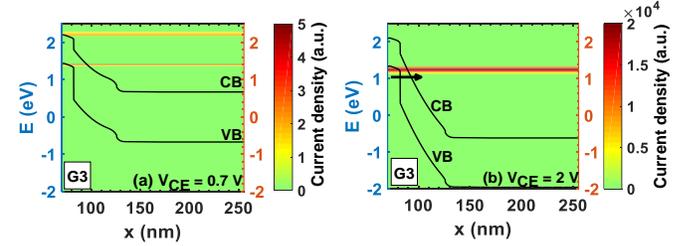

FIG. 10: Spectral current distribution $J_C(x, E)$ for G3 at (a) $V_{CE}$ = 0.7 V and (b) $V_{CE}$ = 2 V calculated in the ballistic limit of transport. The solid lines refer to the conduction and valence bands of the different materials. Red indicates high current concentrations, green no current. The variable x is the distance from the top of the emitter contact. The emitter region has been suppressed to enable a ballistic injection of electrons from the base into the collector.

FIG 8: (a, left) Emitter (x = 0 nm, solid blue line) and collector (x = 700 nm, dashed red line) spectral current distribution of the G0 DHBT at T = 400 K, $V_{CE}$ = 1 V, and $f_{T,peak}$. The results were obtained with model 2 in the presence of electron-phonon scattering. (a, right) Extended spectral current distribution $J_C(x,E)$ over the entire simulation domain. Red indicates high current concentrations, green no current. The solid line represents the conduction band edge of the intrinsic region. The variable x is the distance from the emitter contact. (b) Same as (a), but for G3.

If, instead of assuming $\Delta T$ = 100 K, we explicitly compute this quantity with model 1 or 2, we find that the simulated temperature rise $\Delta T$ induced by self-heating does not exceed 5-10 K, even for G3, where the current density is the highest. In fact, two competing effects compensate each other as the dimension of the device is scaled down. On one side, higher current densities enhance the generation of phonons and boost the backscattering probability, as mentioned earlier. On the other side, shorter base and emitter lengths tend to reduce the phonon emission rate, which is beneficial to the current and cut-off frequency.

The relatively limited impact of electron-phonon scattering is best visible in Fig. 8, where the spectral current distribution at the beginning of the emitter, at the end of the sub-collector, and throughout the device is depicted for the G0 and G3 DHBTs at T = 400 K, $V_{CE}$ = 1 V, and $f_{T,peak}$. From the G0 data, it appears that the electron population loses part of its energy between the emitter and the collector, but that loss remains below 15%. When the total length of collector and sub-collector is reduced from 495 nm in G0 to 172 nm in G3, the energy loss becomes almost negligible, because the distance over which electrons can emit phonons is in the same order of magnitude as that of the mean free path for scattering. Therefore, the spectral current distribution does not vary much between the emitter and collector. Again, it is worth stressing out that the inclusion of lateral dimensions, especially in the G3 device, would probably lift the temperature to higher values by confining the electrons and phonons within an ultra-scaled lateral area [26].

Finally, the accumulated transit times of G3 at $f_{T,peak}$, with and without electron-phonon interactions, at 300 K and 400 K, are presented in Fig. 9. The inclusion of electron-phonon interactions mainly affects the base-collector region, where the slope of the curve slightly increases. Going from 300 to 400 K strengthens the electron-phonon interactions, which has a direct influence on the transit time through the base. Note that the build-up of the transit time from emitter to collector, as described above, exhibits a similar behavior in all DHBT generations. This is why only the results for G3 are presented in Fig. 9.

## IV. BREAKDOWN CHARACTERISTICS

Another essential parameter that must be taken into account when designing future DHBT technologies is their breakdown voltage $BV_{CEO}$. All $BV_{CEO}$ simulations presented here have been carried out in the ballistic limit of transport since the factors that matter are the electrostatic potential (calculated with the HD model) and the band-to-band tunneling (BTBT) current (provided by the QT solver). None of them is strongly impacted by the introduction of electron-phonon interaction in direct-gap semiconductors.

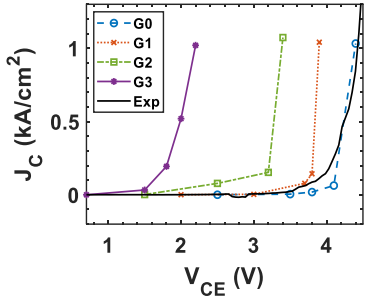

FIG. 11: Common-emitter breakdown characteristics "$J_C$ vs. $V_{CE}$", as computed with the 1-D QT solver in the ballistic limit of transport, keeping only the base-collector junction part of the HD electrostatic potential into account. The solid line without marker represents the experimental measurements for G0.

TABLE I. Simulated performance-related parameters of the type-II InP/GaAsSb DHBTs of generations G0 to G3.

| Parameter | G0 | G1 | G2 | G3 |
|---|---|---|---|---|
| Scaling factor $\gamma$ (%) | 100 | 70 | 50 | 35 |
| $W_E$ (nm) | 300 | 150 | 75 | 37.5 |
| $T_B$ (nm) | 20 | 17 | 14.5 | 12.3 |
| $T_C$ (nm) | 125 | 87.5 | 62.5 | 43.8 |
| $f_{Ts}$ (300 K, model 1) (THz) | 0.46 | 0.76 | 1.14 | 1.66 |
| $f_{Ts}$ (400 K, model 1) (THz) | 0.42 | 0.71 | 1.11 | 1.62 |
| $f_{Ts}$ (400 K, model 2) (THz) | 0.36 | 0.75 | 1.09 | 1.57 |
| $J_{Cs}$ (300 K, model 1) (@$f_{T,peak}$) (mA/µm$^2$) | 8.1 | 9.9 | 29.6 | 49.5 |
| $J_{Cs}$ (400 K, model 1) (@$f_{T,peak}$) (mA/µm$^2$) | 7.7 | 9.5 | 28.7 | 48.2 |
| $J_{Cs}$ (400 K, model 2) (@$f_{T,peak}$) (mA/µm$^2$) | 7.8 | 9.7 | 28.6 | 46.7 |
| $BV_{CEO}$ (V) | 4.4 | 3.9 | 3.4 | 2.2 |

To be able to compute the ballistic BTBT current from the valence band of the base into the conduction band of the collector, the emitter must be cut away so that electrons can be directly injected from the left electrode, as depicted in Fig. 10. The potential profiles are extracted from the 2-D HD model with the base open ($I_B = 0$). By doing so, electrons exhibit a bulk-like energy distribution in the base region instead of being characterized by discrete (quantized) state distribution caused by the short base length. This approximation does not have a strong impact on the results, as demonstrated in Fig. 11, where the $BV_{CEO}$ of all DHBT generations are displayed, together with G0 measurements. Theoretically, the breakdown behavior could be induced by two factors: the tunneling breakdown due to BTBT and the avalanche breakdown due to impact ionization [30]. However, since the tight-binding theory implemented in the proposed TCAD framework is essentially a single-electron model, it is not capable of capturing the latter effect induced by many-body electron-electron interactions [31]. As a result, only the contribution coming from BTBT is taken into account. Considering the fact that the calibrated HD+QT approach very accurately reproduces the experimental G0 behavior, where a $BV_{CEO}$ value of 4.4 V can be deduced, when extracted at $J_C = 1$ kA/cm$^2$. We could expect that BTBT current occurring at high $V_{CE}$ is the dominating factor in this phenomenon for the type-II DHBTs due to the relatively elevated electric field. This assumption is further justified by the measured temperature dependence of $I_C$ presented in [32]: a positive temperature coefficient is reported, which indicates that the breakdown is governed by band-to-band tunneling currents [33]. Consequently, the same approach is applied to G1-G3.

It can be further seen in Fig. 11 that, as the device dimensions are scaled down, $BV_{CEO}$ shifts towards lower voltages because of the increasing electric field, which favors band-to-band tunneling [34]. In G0, G1, and G2, BTBT remains manageable as $BV_{CEO}$ does not shrink below 3 V. However, in G3, the BTBT current drastically increases, lowering $BV_{CEO}$ to 2.2 V. The early turn-on of the BTBT in G3 is highlighted in Fig. 10 by comparing the spectral currents at $V_{CE} = 0.7$ and 2 V. The ultimately scaled device dimensions combined with higher doping concentrations allow BTBT to manifest itself at low $V_{CE}$ and to cause a breakdown already at $V_{CE} = 2.2$ V.

V. CONCLUSION

The intrinsic performance of aggressively scaled type-II InP/GaAsSb DHBTs towards and beyond 1 THz has been simulated using a calibrated multi-scale TCAD framework.

Starting from an existing epitaxial structure, its vertical and lateral dimensions have been reduced according to the scaling prescriptions of [8], producing three generations of transistors. By combining 2-D HD and 1-D QT simulations, the DC and AC characteristics of these devices have been investigated, in the ballistic limit of transport and in the presence of electron-phonon interactions, using two different scattering models. Reductions in both collector current densities and transit frequencies are observed when the electron-phonon interaction is turned on. At the same time, a noticeable, but moderate increase of the lattice temperature ($< 10$ K) across the center of the structures is found for all device generations. Finally, the breakdown characteristics of the transistors have been analyzed.

Table I summarizes the relevant AC and DC figures-of-merit of all simulated type-II InP/GaAsSb DHBT generations. Simulations predict that a peak $f_T$ higher than 1 THz could be possibly reached for both G2 and G3, even if electron-phonon interactions are taken into account. For each generation, several technology challenges have to be addressed, from the scaling of the device dimensions (layer thicknesses and lateral extensions), to the increase of the doping concentrations and the reduction of the parasitic elements, in particular the junction capacitances. Considering the time that is necessary to overcome the possible fabrication bottlenecks, it can be estimated that the G1 device generation can be realized in about 6 months, since process for fabricating 150-nm wide emitter already exists. The additional work that is required is the wet etching calibrations for new thickness of the layers. However, in terms of G2 and G3, reliable process for fabricating contact metals and efficient surface treatment technology have to be developed. Therefore, we estimate that G2 and G3 devices can be realized in 1~1.5 years, and in 2~2.5 years, respectively. While the higher $f_T$ is obtained for G3 (1.66 THz at 300 K, ~1.6

THz at 400 K), its breakdown voltage of 2.2 V appears to be too low for practical applications, as it limits the dynamic range of operation of the DHBT [35, 36]. G2 might therefore be a more suitable target. Ultimately, compromises will be needed between ultra-scaled dimensions and breakdown voltages.

In terms of modeling, this work presented an accurate and reliable simulation framework to shed light on the intrinsic performance of type-II DHBTs with complex epitaxial structures. It captures their bandstructure, quantum mechanical, and transient effects, both in the ballistic limit of transport and with electron-phonon scattering. As a future step, the influence of parasitic elements should be studied as a function of the reduced device dimensions. Such investigations will require going beyond 1-D QT simulations.

ACKNOWLEDGMENT

This work was supported in part by the Swiss National Science Foundation (SNF) under Grant 200021L_169413 and by the Agence Nationale de la Recherche (ANR) in the frame of the project ULTIMATE under Grant ANR-16-CE93-0007. The use of computational resources from the Swiss National Supercomputing Center (CSCS) under project s876 is acknowledged.

APPENDIX A: MODEL 1

To investigate the increase of lattice temperature in *model 1*, electron and phonon transport are coupled to each other through scattering self-energies within the NEGF formalism. As compared to pure electronic transport, the dynamical matrix of the G0 to G3 structures must be created to model the displacement of phonons. To minimize the computational burden, the simulation domain is restricted to a 1-D line, as for electrons [9].

The valence-force-field (VFF) method, which is the pendant of tight-binding for phonons, is used to construct all dynamical matrices. As a first approximation, the phonon properties are assumed to be the same throughout the device because the phonon bandstructures of the involved III-V semiconductors do not vary much from one compound to the other. The VFF model and parameters of [37] are employed, without longitudinal-transverse optical phonon splitting.

The scattering self-energies coupling the electron and phonon populations have the following form [10]:

$$\Sigma^{\gtrless}(E,k) = i\lambda \sum_q \int \frac{\partial \hbar\omega}{2\pi} \nabla H \cdot G^{\gtrless}(E - \hbar\omega, k - q) \cdot \nabla H \cdot D^{\gtrless}(\omega, q), \quad (A1)$$

$$\Pi^{\gtrless}(\omega,q) = -i\lambda \sum_k \int \frac{\partial E}{2\pi} \nabla H \cdot G^{\gtrless}(E - \hbar\omega, k + q) \cdot \nabla H \cdot G^{\lessgtr}(E, k). \quad (A2)$$

In these equations, $\Sigma^{\gtrless}$ is the lesser (<) or greater (>) electron-phonon scattering self-energy, $\Pi^{\gtrless}$ its phonon-electron counterpart, $G^{\gtrless}$ the lesser/greater electron Green's function, while $D^{\gtrless}$ is the lesser/greater phonon Green's function. All these quantities depend on the electron energy (momentum) $E$ ($k$) or phonon frequency (momentum) $\omega$ ($q$). Key ingredients of Eqs. (A1) and (A2) are the $\nabla H$ blocks, which represent the derivative of the Hamiltonian matrix along the bonds connecting two nearest-neighbor atoms. Those elements are calculated from the selected tight-binding parameters, where strain is added according to the model of [38].

The scattering self-energies are computed self-consistently with the electron and phonon Green's Functions within the so-called Born approximation, which ensures current and energy conservation. In order to reproduce the "$f_T$ vs. $I_C$" characteristics of G0 in Fig. 4, both $\Sigma^{\gtrless}(E,k)$ and $\Pi^{\gtrless}(\omega, q)$ are scaled by the same factor $\lambda$, which remains the same for the entire range of injected currents and for all DHBT generations. This scaling is justified by the fact that the electron-phonon scattering self-energy matrices are assumed to be diagonal and the phonon-electron ones only account for the coupling with nearest-neighbor atoms [10]. By scaling these scattering self-energies, the missing inter-atomic interactions can be partly compensated [39]. It should be noted that the influence of any other mechanism leading to a current reduction, e.g. additional scattering sources or unidentified parasitic elements, might be unwillingly cast into the $\lambda$ scaling factor during the fitting procedure. We do not expect this mixture of multiple effects to significantly affect our results.

Finally, we would like to draw the reader's attention on the fact that the electron-phonon interactions in Eqs. (A1) and (A2) rely on the deformation potential theory. Using these expressions, polar optical phonons (POP), which are dominant in III-V semiconductors [40], cannot be taken into account. For example, the widely used approach based on the Fröhlich coupling element indirectly includes the phonon Green's function $D^{\gtrless}$, thus preventing a self-consistent coupling between the electron and phonon populations.

APPENDIX B: MODEL 2

To incorporate POP-like interactions in our simulations, we have developed an alternative electron-phonon scattering approach, *model 2*, that uses the following, relatively simple equation for the $\Sigma^{\gtrless}$ self-energy

$$\Sigma^{\gtrless}(E,k) = D_{e-ph}^2 \left( n_\omega G^{\gtrless}(E + \hbar\omega, k) + (n_\omega + 1)G^{\gtrless}(E - \hbar\omega, k) \right). \quad (B1)$$

Here, $D_{e-ph}$ is the electron-phonon scattering strength and $n_\omega = 1/(\exp(\hbar\omega/k_B T) - 1)$ the Bose-Einstein distribution function for phonons with frequency $\omega$, at temperature $T$. To compute $T$, it is assumed that phonons are at local equilibrium within each atomic unit cell of the simulation domain, i.e. they obey Bose-Einstein distribution functions with local temperature values. The 1-D classical Fourier heat equation can then be recalled to obtain $T$

$$\frac{d}{dx}\kappa_{th}(x)\frac{d}{dx}T(x) = -Q(x), \quad (B2)$$

where $\kappa_{th}(x)$ is the thermal conductivity at $x$, while $Q(x)$ is the amount of dissipated power per volume (unit: W/m$^3$) at the same location. It can be calculated from the electrical energy current $J_{dE}$ (unit: W/m$^2$) that flows through the DHBT structure

$$J_{dE}(x_i) = \frac{q}{\hbar L_y L_z} \int \frac{dE}{2\pi} (E - E_F) tr(H_{ii+1} \cdot G_{i+1i}^{<} - G_{ii+1}^{<} \cdot H_{i+1i}), \quad (B3)$$

$$Q(x_i) = -\frac{dJ_{dE}(x)}{dx}\bigg|_{x_i} \approx -\frac{J_{dE}(x_{i+1}) - J_{dE}(x_i)}{L_x}. \quad (B4)$$

In these equations, the $L_i$'s are the dimensions of the atomic unit cells along all Cartesian coordinates $i \in \{x, y, z\}$, $E$ is the electron energy, and $E_F$ the Fermi level of the emitter. The $H_{ii+1}$ Hamiltonian matrix blocks connect a unit cell situated at $x = x_i$ to another one at $x = x_{i+1}$.

Each simulation begins with the self-consistent solution of Eq. (B1) and the electron Green's Functions under the assumption that the temperature is constant over the entire device structure. Next, $J_{dE}(x_i)$ and $Q(x_i)$ are evaluated for each unit cell and plugged into Eq. (B2) to produce the temperature profile $T(x_i)$. As boundary conditions, we set $T(x_1)$ and $T(x_{N_x})$ to pre-defined values, generally the temperature of the environment. Finally, $T(x_i)$ is inserted into the Bose-Einstein distribution function, $n_\omega$, which becomes position-dependent. The procedure starts again from the $G^{\gtrless}$-$\Sigma^{\gtrless}$ loop, but with $n_\omega(T(x_i))$, and continues until the temperature does not vary any more between two consecutive iterations.

All parameters used in this work are summarized in Table II. The scattering intensity, $D_{e-ph}$, was adjusted to best reproduce the measured $f_T$ of G0 over the whole range of current injection levels (see Fig. 4). The thermal conductivities of InGaAs and GaAsSb are chosen to be small (5 W/mK) because phonons are extremely sensitive to (alloy) disorder. The $\kappa_{th}$ of InP is slightly decreased (50 W/mK) as compared to its bulk value (68 W/mK) to account for surface effects. Note that the same phonon energy (30 meV) is considered along the entire device structure, regardless of the material (InP, InGaAs, GaAsSb). The inclusion of InGaAs is justified by the fact that part of the emitter contact of the DHBT is made of this material, as described in [9].

TABLE II. List of material parameters for model 2.

| $D_{e-ph}$ (meV) | $\hbar\omega$ (meV) | $\kappa_{th,InP}$ (W/mK) | $\kappa_{th,InGaAs}$ (W/mK) | $\kappa_{th,GaAsSb}$ (W/mK) |
|---|---|---|---|---|
| 50 | 30 | 50 | 5 | 5 |

DATA AVAILABILITY STATEMENT

The data that support the findings of this study are available from the corresponding author upon reasonable request.

REFERENCES

[1] N. Kashio, K. Kurishima, M. Ida, and H. Matsuzaki, "Over 450-GHz ft and fmax InP/InGaAs DHBTs with a Passivation Ledge Fabricated by Utilizing SiN/SiO2 Sidewall Spacers," IEEE. Trans. Electron Devices 61, 3423 (2014).
[2] W. Quan, A. M. Arabhavi, R. Flückiger, O. Ostinelli, and C. R. Bolognesi, "Quaternary graded-base InP/GaInAsSb DHBTs with $f_T/f_{MAX}$ = 547/784 GHz," IEEE Electron Dev. Lett. 39, 1141 (2018).
[3] A. Scavennec, M. Sokolich, and Y. Baeyens, "Semiconductor technologies for higher frequencies," IEEE Microwave Magazine 10, 77 (2009).
[4] C. R. Bolognesi, W. Quan, A. M. Arabhavi, T. Saranovac, R. Flückiger, O. Ostinelli, X. Wen, and M. Luisier, "Advances in InP/Ga(In)AsSb double heterojunction bipolar transistors (DHBTs)," Jpn. Journal of Appl. Phys. 58, SB0802 (2019).
[5] M. Schröter, G. Wedel, B. Heinemann, C. Jungemann, J. Krause, P. Chevalier, and A. Chantre, "Physical and electrical performance limits of high-speed SiGeC HBTs—part I: vertical scaling," IEEE. Trans. Electron Devices 58, 3687 (2011).
[6] M. Schröter, T. Rosenbaum, P. Chevalier, B. Heinemann, S. P. Voinigescu, E. Preisler, J. Böck, and A. Mukherjee, "SiGe HBT technology: future trends and TCAD-based roadmap," Proc. IEEE 105, 1068 (2017).
[7] S.-M. Hong and C. Jungemann, "A fully coupled scheme for a Boltzmann–Poisson equation solver based on a spherical harmonics expansion," J. Comput. Electron. 8, 225 (2009).
[8] M. J. W. Rodwell, M. Urteaga, T. Mathew, D. Scott, D. Mensa, Q. Lee, J. Guthrie, Y. Betser, S. C. Martin, R. P. Smith, S. Jaganathan, S. Krishnan, S. I. Long, R. Pullela, B. Agarwal, U. Bhattacharya, L. Samoska, and M. Dahlström, "Submicron scaling of HBTs," IEEE. Trans. Electron Devices 48, 2606 (2001).
[9] X. Wen, C. Mukherjee, C. Raya, B. Ardouin, M. Deng, S. Frégonèse, V. Nodjiadjim, M. Riet, W. Quan, A. M. Arabhavi, O. Ostinelli, C. R. Bolognesi, T. Zimmer, C. Maneux, and M. Luisier, "A multiscale TCAD approach for the simulation of InP DHBTs and the extraction of their transit times," IEEE. Trans. Electron Devices 66, 5084 (2019).
[10] R. Rhyner and M. Luisier, "Atomistic modeling of coupled electron-phonon transport in nanowire transistors," Phys. Rev. B 89, 235311 (2014).
[11] M. Bescond, D. Logoteta, F. Michelini, N. Cavassilas, T. Yan, A. Yangui, M. Lannoo, and K. Hirakawa, "Thermionic cooling devices based on resonant-tunneling AlGaAs/GaAs heterostructure," J. Phys.: Condens. Matter 30, 064005 (2018).
[12] *Sentaurus Device User Guide Version 2019.12* (Synopsys Inc., CA, USA, 2019).
[13] M. Luisier, A. Schenk, W. Fichtner, and G. Klimeck, "Atomistic simulation of nanowires in the sp3d5s∗ tight-binding formalism: from boundary conditions to strain calculations," Phys. Rev. B 74, 205 (2006).
[14] B. Heinemann, H. Rücker, R. Barth, F. Bärwolf, J. Drews, G. G. Fischer, A. Fox, O. Fursenko, T. Grabolla, F. Herzel, J. Katzer, J. Korn, J. Krüger, P. Kulse, T. Lenke, M. Lisker, S. Marschmeyer, A. Scheit, D. Schmidt, J. Schmidt, M. A. Schubert, A. Trusch, C. Wipf, and D. Wolansky, "SiGe HBT with fx/fmax of 505 GHz/720 GHz," 2016 IEEE International Electron Devices Meeting (IEDM) (2016).
[15] Z. Griffith, E. Lind, M. J. W. Rodwell, X.-M. Fang, D. Loubychev, Y. Wu, J. M. Fastenau, A. W. K. Liu, "Sub-300 nm InGaAs/InP Type-I DHBTs with a 150 nm collector, 30 nm base demonstrating 755 GHz fmax and 416 GHz fT," 2007 IEEE 19th International Conference on Indium Phosphide & Related Materials (2007).
[16] I. Vurgaftman, J. R. Meyer, and L. R. Ram-Mohan, "Band parameters for III–V compound semiconductors and their alloys," Journal of Appl. Phys. 89, 5815 (2001).
[17] J. J. H. van den Biesen, "A simple regional analysis of transient times in bipolar transistors," Solid-State Electron 29, 529 (1986).
[18] U. K. Mishra, and J. Singh, Semiconductor device physics and design. Springer, Dordrecht, 2008.
[19] M. Alexandrova, "Development and optimization of high-speed InP/GaAsSb double heterojunction bipolar transistors", PhD Thesis, ETH Zürich, Zürich, Switzerland, 2015.
[20] C. Mukherjee, M. Couret, V. Nodjiadjim, M. Riet, J. –Y. Dupuy, S. Frégonèse, T. Zimmer, and C. Maneux, "Scalable Modeling of Thermal Impedance in InP DHBTs Targeting Terahertz Applications," IEEE Trans. Electron Devices 66, 2125 (2019).
[21] B. Grandchamp, V. Nodjiadjim, M. Zaknoune, G. A. Koné, C. Hainaut, J. Godin, M. Riet, T. Zimmer, and C. Maneux, "Trends in Submicrometer InP-Based HBT Architecture Targeting Thermal Management", IEEE Trans. Electron Devices 58, 2566 (2011).
[22] P. Y. Yu, and M. Cardona, *Fundamentals of Semiconductors*, (Springer-Verlag, Berlin, 2010) p.203-241.
[23] M. Luisier, and G. Klimeck, "Atomistic full-band simulations of silicon nanowire transistors: effects of electron-phonon scattering," Physical Rev. B 80, 155430 (2009).
[24] H. Kamrani, T. Kochubey, D. Jabs, and C. Jungemann, "Electrothermal simulation of SiGe HBTs and investigation of experimental extraction methods for junction temperature," Proc. IEEE Int. Conf. Simulation Semiconductor Process. Devices (SISPAD), 108 (2015).
[25] L. Harrison, M. Dahlstrom, S. Krishnan, Z. Griffith, Y. M. Kim, and M. J. W. Rodwell, "Thermal Limitations of InP HBTs in 80- and 160-Gb ICs," IEEE. Trans. Electron Devices 51, 529 (2004).
[26] M. J. W. Rodwell, M. Le, and B. Brar, "InP Bipolar ICs: Scaling Roadmaps, Frequency Limits, Manufacturable Technologies," Proceedings of the IEEE 96, 271 (2008).
[27] J. C. Li, T. Hussain, D. A. Hitko, Y. Royter, C. H. Fields, I. Milosavljevic, S. Thomas, R. D. Rajavel, P. M. Asbeck, and M. Sokolich, "Reduced temperature S-parameter measurements of 400+GHz sub-micron InP DHBTs," Solid-State Electronics 51, 870 (2007).
[28] W. Liu, Hin-Fai Chau and E. Beam, "Thermal properties and thermal instabilities of InP-based heterojunction bipolar transistors," IEEE Transactions on Electron Devices 43, 388 (1996).
[29] R. Flückiger, "Monolithic Microwave Integrated Circuits Based on InP/GaAsSb Double Heterojunction Bipolar Transistors", PhD Thesis, ETH Zürich, Zürich, Switzerland, 2015.

[30] M. S. Tyagi, "Zener and avalanche breakdown in silicon alloyed p-n junctions—I: Analysis of reverse characteristics," Solid-State Electronics 11, 99 (1968).
[31] E. Nielsen, R. Rahman, and R. P. Muller, "A many-electron tight binding method for the analysis of quantum dot systems," Journal of Applied Physics 112, 114304 (2012).
[32] C. Mukherjee, C. Raya, B. Ardouin, M. Deng, S. Frégonèse, T. Zimmer, V. Nodjiadjim, M. Riet, J.-Y. Dupuy, M. Luisier, Q. Quan, A. Arabhavi, C. R. Bolognesi, and C. Maneux, "Scalable Compact Modeling of III–V DHBTs: Prospective Figures of Merit Toward Terahertz Operation," IEEE Transactions on Electron Devices 65, 5357 (2018).
[33] H. Czichos, T. Saito T, and L. Smith, Springer handbook of metrology and testing, (Springer, New York, 2011).
[34] E. O. Kane, "Theory of tunneling," Journal of App. Phys. 32, 83 (1961).
[35] H. Xu, "High-speed type-II GaAsSb/InP DHBTs for mixed-signal IC applications", PhD Thesis, University of Illinois at Urbana-Champaign, Urbana, Illinois, 2014.
[36] L. E. Larson, "Silicon technology tradeoffs for radio-frequency/mixed-signal "systems-on-a-chip"," IEEE Transactions on Electron Devices 50, 683 (2003)
[37] K. Vuttivorakulchai, "Thermoelectricity of Nanowires", PhD Thesis, ETH Zürich, Zürich, Switzerland, 2019.
[38] T. B. Boykin, G. Klimeck, R. C. Bowen, and F. Oyafuso, "Diagonal parameter shifts due to nearest-neighbor displacements in empirical tight-binding theory", Phys. Rev. B 66, 125207 (2002).
[39] R. Rhyner and M. Luisier, "Phonon-limited low-field mobility in silicon: Quantum transport vs. linearized Boltzmann Transport Equation", J. Appl. Phys. 114, 223708 (2013).
[40] R. Lake, G. Klimeck, R. C. Bowen, and D. Jovanovic, "Single and multiband modeling of quantum electron transport through layered semiconductor devices," Journal of Appl. Phys. 81, 7845 (1997).